\documentclass[aps,prb,twocolumn,color,superscriptaddress,floatfix,nofootinbib]{revtex4-2}
\usepackage[english]{babel} 
\usepackage{amssymb}
\usepackage{amsmath}
\usepackage{mathtools}
\usepackage{txfonts}
\usepackage{mathdots}
\usepackage[normalem]{ulem}
\usepackage[dvips]{graphicx}
\usepackage{epsfig}
\usepackage{graphicx}
\usepackage{array}
\usepackage{amssymb}
\usepackage{amsfonts}
\usepackage{amsmath}
\usepackage{mathrsfs}
\usepackage{booktabs}
\usepackage{threeparttable}
\usepackage{multirow}
\usepackage{subfigure}
\usepackage{epsfig}
\usepackage{threeparttable}
\usepackage{chngpage}
\usepackage{float}
\usepackage{xcolor}
\usepackage{bm}
\usepackage{graphicx}
\usepackage[toc]{appendix}
\usepackage{tabularx}
\usepackage{graphicx}
\usepackage{adjustbox}
\usepackage{comment}
\usepackage{hyperref}

\usepackage{tabularx}
\usepackage{graphicx}
\usepackage{adjustbox}

\hypersetup{
    unicode=false,     
    pdftoolbar=false,  
    pdfmenubar=true,   
    pdffitwindow=false, 
    pdfstartview={FitH},
    pdftitle={},    
    pdfauthor={Authors},     
    pdfsubject={},   
    pdfcreator={},   
    pdfproducer={}, 
    pdfkeywords={quantum many-body scars} {superconducting processor} {quantum state tomography}, 
    pdfnewwindow=true,
    colorlinks=true,
    linkcolor=black,
    citecolor=blue, 
    filecolor=magenta,
    urlcolor=blue
}

\begin{document}

\title{Origin of Hilbert space quantum scars in unconstrained  models  }

\author{Zexian Guo} 
\thanks{These authors contributed equally}

\author{Bobo Liu} 
\thanks{These authors contributed equally}

\author{Yu Gao} 
\author{Ang Yang}
\author{Junlin Wang} 
\author{Jinlou Ma} 

\author{Lei Ying} 
\email{leiying@zju.edu.cn}
\affiliation{School of Physics and Interdisciplinary Center for Quantum Information, Zhejiang University, Hangzhou 310027, China}

\date{\today}

\begin{abstract}
Quantum many-body scar is a recently discovered phenomenon weakly violating eigenstate thermalization hypothesis, and it has been extensively studied across various models.
However, experimental realizations are mainly based on constrained models such as the $PXP$ model. Inspired by recent experimental observations on the superconducting platform in Refs.~[Nat. Phys. 19, 120 (2022)] and [arXiv:2211.05803], we study a distinct class of quantum many-body scars based on a half-filling hard-core Bose-Hubbard model, which  is generic to describe in many experimental platforms. It is the so-called Hilbert space quantum scar as it originates from a subspace with a hypercube geometry weakly connecting to other thermalization regions in Hilbert space. Within the hypercube, a pair of collective Fock states do not directly connect to the thermalization region, resulting in slow thermalization dynamics with remarkable fidelity revivals with distinct differences from dynamics of other initial states. This mechanism is generic in various real-space lattice configurations, including one-dimensional Su-Schrieffer-Heeger chain, comb lattice, and even random dimer clusters consisting of dimers. In addition, we develop a toy model based on Hilbert hypercube decay approximation, to explain the spectrum overlap between the collective states and all eigenstates. Furthermore, we explore the Hilbert space quantum scar in two- and three-dimensional Su-Schrieffer-Heeger many-body systems, consisting of tetramers or octamers, respectively. This study makes quantum many-body scar state more realistic in applications such as quantum sensing and quantum metrology.
\end{abstract}

\maketitle

\section{Introduction}
Violations of eigenstate thermalization hypothesis (ETH) in strongly correlated quantum many-body systems are rare~\cite{Deutsch:1991,Srednicki:1994,kim2014testing}, but quite important due to their potential applications in quantum information, including quantum integrability~\cite{etingof1995quantum,prosen1998time}, many-body localization~\cite{ponte2015many,schindler2017probing,nandkishore2015many,serbyn2014interferometric}, and Hilbert space fragmentation~\cite{de2019dynamics,regnault2022quantum}. These phenomena all strongly break the ergodicity of many-body systems.  Recently, a new type of coherent many-body states has attracted great interest as some special eigenstates, embedded in the thermalized spectrum, exhibit non-ergodic behavior. Thus,  being analogous to scars in single-particle quantum billiards, it was named quantum many-body scar (QMBS), which was first discovered in a Rydberg atom array in strong nearest-neighbor interaction limit~\cite{BSKL:2017}, reduced to the kinetically constrained $PXP$ model~\cite{TMAS:2018,SVLSDG:2021}. 

Quantum many-body scar (QMBS) states have been studied in numerous quantum many-body models, including the Heisenberg spin-$1$ $XY$ model~\cite{SI:2019,SYK:2020}, the Affleck-Kennedy-Lieb-Tasaki model~\cite{MRB:2018,Shiraishi:2019}, the extended Hubbard model~\cite{MM:2018,MRB:2020,DHTP:2021,scherg2021observing}, the Ising model~\cite{VMS:2020}, frustrated~\cite{MHSR:2020,LMPC:2020} and topological~\cite{OCMC:2019,WSSN:2021} lattices, quantum Hall systems~\cite{MB:2020,NWY:2020}, Floquet-driven systems~\cite{MNSS:2020,MTK:2020}, systems with a flat band~\cite{KMH:2020,HDGC:2020,MHSR:2020}, and two-dimensional systems~\cite{LCH:2020}. Besides, several general theoretical frameworks have been developed for QMBS states, including the embedding method~\cite{SM:2017,BMP:2019} and quasi-symmetry groups~\cite{RLF:2021}. QMBS states were used for generating Greenberger-Horne-Zeilinger (GHZ) entanglement states~\cite{OLKS:2019} and for robust quantum sensing~\cite{Dooley:2021, dooley2023entanglement, desaules2022extensive}. Due to their ability to defy thermalization, QMBS states have also been proposed for manipulating and storing quantum information~\cite{khemani2019local,SAP:2021}.

Nevertheless, experimental realizations are quite limited, mainly on platforms such as Rydberg atoms~\cite{BSKL:2017,TMAS:2018,SVLSDG:2021}, ultracold $^{87}$Rb atoms~\cite{Su2022}, and digital quantum simulator based on the $PXP$ model~\cite{PhysRevResearch.4.043027} and Li atoms based on the spin-1/2 XXZ model~\cite{Jepsen2021}. Quite recently, a new class of QMBS was found in superconducting qubits~\cite{zhang2022many} in a one-dimensional (1D) Su-Schrieffer–Heege (SSH) model with irregular cross couplings based on the spin-1/2 XXZ model, where a pair of special collective Fock states have revivals in fidelity dynamics with a stable period. This class of QMBS is called Hilbert space quantum scar, having a distinct origin from the constrained models in Hilbert space. Compared to most other QMBS states, we find this kind of scarred states is experimental-friendly to be prepared in a superconducting processor or other simulation platforms, which are based on unconstrained models. Also, it possesses slow entanglement entropy growth, which can be applied to the generation of GHZ states~\cite{OLKS:2019} and quantum metrology~\cite{Dooley:2021, dooley2023entanglement, desaules2022extensive}.

In this paper, we systematically study the origin of the Hilbert space quantum scar in-depth. This kind of scar states in the hard-core Bose-Hubbard model had not been theoretically investigated. Thus, in detail,  we study its mechanism from the viewpoint in Hilbert space in Sec.~\ref{sec:model}, in which we demonstrate the origin of this class of QMBS and propose a decay approximation of the Hilbert hypercube to describe the scarring mechanism. Furthermore, in Sec.~\ref{sec:dimer_num}, we numerically verify the scarring phenomenon from both the eigenstates and the special Fock state dynamics for the 1D SSH chain, comb lattice, and random dimer cluster. In Sec.~\ref{sec:2d}, we expand the Hilbert space quantum scar to higher dimensions, whose SSH lattice is consisting of a set of tetramers in two dimensions, where more kinds of special Fock states in higher dimensions have slow thermalized dynamics.

\section{Models and Hilbert hypercube }\label{sec:model}

\subsection{Models and Hamiltonians}

At first, we consider the Bose-Hubbard model, which is a generic model for various quantum electrodynamics (QED) systems. The generic Hamiltonian is given by $(\hbar=1)$
\begin{equation}
\begin{split}
	\hat{H}_\mathrm{BH} = \sum_{m} \omega_m\hat{\sigma}^+_m\hat{\sigma}^-_m
	+& \sum_{m,n} J_{mn}  \left( \hat{\sigma}^+_m\hat{\sigma}^-_n + \mathrm{h.c.}  \right) \\
	+& \sum_{m}  U_m\hat{\sigma}^+_m\hat{\sigma}^+_m\hat{\sigma}^-_m\hat{\sigma}^-_m ,
\end{split}
\label{eq:hamiltionian_BH}
\end{equation}
where $\hat{\sigma}^+_m\ (\hat{\sigma}^-_m)$ is the raising (lowering) operator of $m$th quantum two-level system (qubit). As the nonlinearity $U_m$ is much larger than the couplings ($U_m/J_{mn} \gg 1$),  Hamiltonian~(\ref{eq:hamiltionian_BH}) can be reduced to the Heisenberg spin-$1/2$ $XY$  Hamiltonian (also known as the hard-core Bose-Hubbard model)
\begin{equation}
	\hat{H}_\mathrm{XY} = \sum_{m} \frac{\omega_m}{2}(\hat{\sigma}^z_m +  \hat{\sigma}_m^0)
	+ \sum_{m,n} \frac{J_{mn}}{2}  \left( \hat{\sigma}^x_m \hat{\sigma}^x_n + \hat{\sigma}^y_m\hat{\sigma}^y_n  \right),  
\label{eq:hamiltionian_XY}
\end{equation}
where $\hat{\sigma}^{x,y,z}_m$ are the spin-$1/2$ Pauli matrices and $\hat{\sigma}^0$ is the identity matrix. We have $\hat{\sigma}^{\pm}_m = 1/\sqrt{2}(\hat{\sigma}^x_m \pm \hat{\sigma}_m^y)$. 
Hamiltonian Eq.~(\ref{eq:hamiltionian_XY}) is generally studied in magneto materials \cite{tsvelik1990field,white1996dimerization,carollo2005geometric,steinigeweg2013eigenstate}, 
circuit-QED~\cite{blais2007quantum,blais2020quantum,clerk2020hybrid,petersson2012circuit}, and waveguide/cavity-QED systems~\cite{koch2010time,vaidya2018tunable,zanner2022coherent}. 
Thus, in this paper, we focus on Hamiltonian~(\ref{eq:hamiltionian_XY}).

Now, we start to study the mechanism of the scarring phenomenon in the dimer cluster system with a Hamiltonian as
\begin{equation}
\hat{H}_0 = \sum_\alpha  \hat{h}_\alpha + \sum_{\alpha, \beta} \hat{h}_{\alpha\beta},
\end{equation}
where $\alpha,\beta=1\cdots N$ are the indices of dimers. We use ``$a,b$'' to denote the sites in a dimer and ``$\hat{a},\hat{b}$'' to represent the operators of sub-dimer sites. 
Then, the single dimer Hamiltonian and intra-dimer Hamiltonian are respectively written as
\begin{equation}
\begin{split}
\hat{h}_\alpha = &  \omega_\alpha \left(\hat{a}^+_\alpha\hat{a}^-_\alpha +\hat{b}^+_\alpha\hat{b}^-_\alpha \right) 
+ J_0 \left(\hat{a}^+_\alpha \hat{b}^-_\alpha +\mathrm{h.c.} \right), \\
\hat{h}_{\alpha\beta} = &
 J_1 \left( \delta \hat{a}^+_\alpha \hat{b}^-_\beta + \delta^\prime \hat{b}^+_\alpha \hat{a}^-_\beta + \delta^{\prime\prime} \hat{a}^+_\alpha \hat{a}^-_\beta
 +\delta^{\prime\prime\prime} \hat{b}^+_\alpha \hat{b}^-_\beta+\mathrm{h.c.} \right).
\end{split}
\label{eq:dimer_hamiltonian}
\end{equation}

At first, we focus on a single dimer with no inter-dimer coupling
(i.e. $J_1=0$), which is crucial to understand these special scarred eigenstates. Two eigenstates with particle-number conservation are written as singlet state $\vert d_\mathrm{s}\rangle=(\vert 10\rangle - \vert 01\rangle)/\sqrt{2}$ with an energy of $-J_0$ and a triplet state $\vert d_\mathrm{t}\rangle=(\vert 10\rangle + \vert 01\rangle)/\sqrt{2}$ with energy of $J_0$. Thus,  the decoupled Hamiltonian of $N$ one-particle dimers ($J_1=0$) can be written as
\begin{equation}
\hat{H}_{\mathrm{eff},J_1=0}=J_0\sum_{\alpha=1}^N \hat{X}_\alpha,
\label{eq:decouple_hamiltonian}
\end{equation}
where $\hat{X}_\alpha= \vert01\rangle_{\alpha}\langle10\vert_{\alpha} + \vert10\rangle_{\alpha}\langle01\vert_{\alpha}$ is the x-component Pauli matrix for the one-particle dimer $\alpha$. It is a famous model considered to hold perfect quantum state transfer and thus exact fidelity revival, no matter what product state we choose to set as an initial states. From analytical calculations, the eigenenergy levels of decoupled Hamiltonian can be written as $E = rJ_0$ with $r = -N, -N+2, ..., N-2, N$, which are equally distributed. 

As the intra-dimer coupling increases, the equal spacing of eigenstates $\Delta E$ becomes large induced by the strongly correlated particles. In general, $\Delta E/J_0 \approx \lambda (J_1/J_0)^2 + 2$, where the parameter $\lambda$ is related to the system size and approaches a constant at the thermodynamic limit. For 1D SSH chain and comb lattice, we show the numerical fitting results in Appendix.~\Ref{App:delta E}.

\subsection{Robust hypercube in Hilbert space}

\begin{figure}
\includegraphics[width=\linewidth]{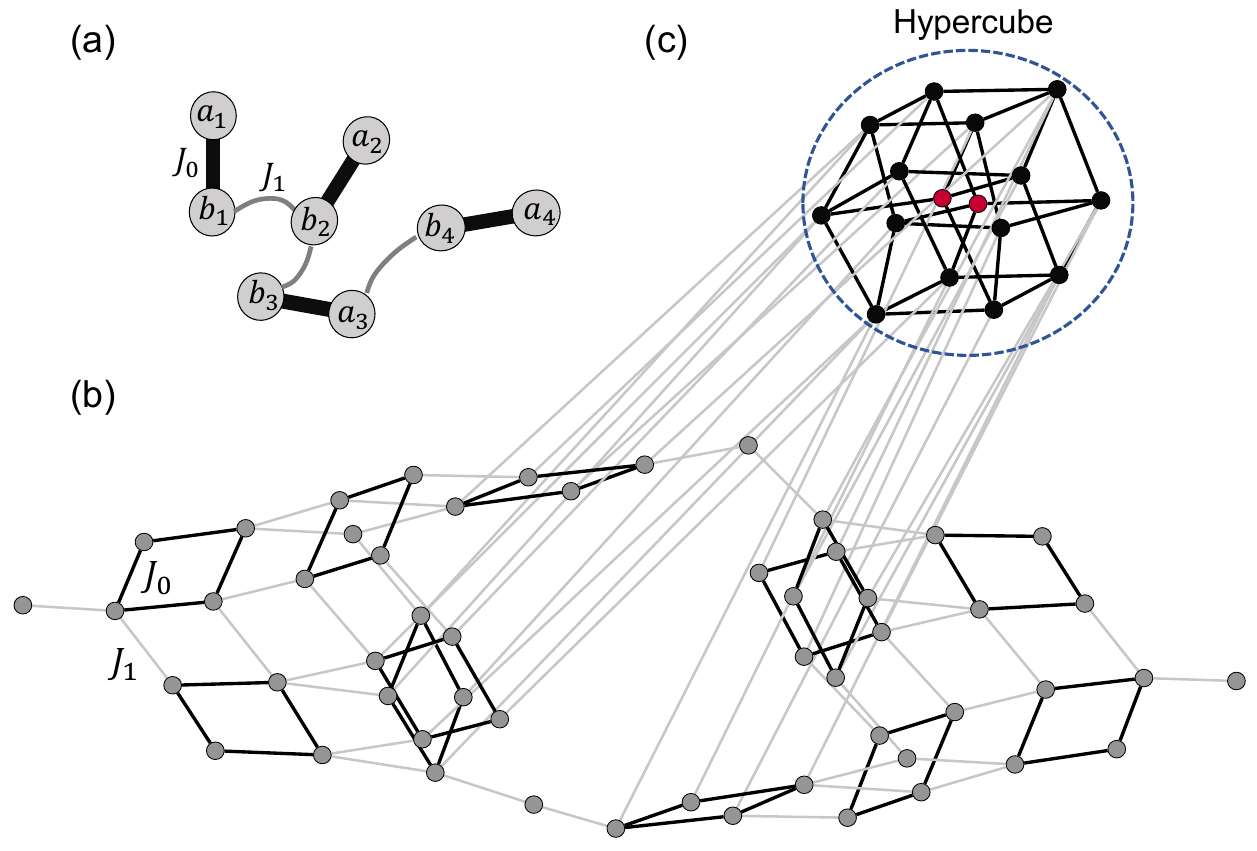}
\caption{
(a) An example of the many-body system consisting of four dimers. (b) and (c) show the adjacency graph in Hilbert space corresponds to the real-space many-body system in (a).  The whole system is divided into hypercube and thermal environment, which are in and outside the blue dashed circle, respectively.  Two scar states (red bullets) are $C=\vert 10101001\rangle$ and $C^\prime=\vert 01010110\rangle$. Bullets belonging to the hypercube denote Fock states $\vert z_{a_1}z_{b_1}z_{a_2}z_{b_2}z_{a_3}z_{b_3}z_{a_4}z_{b_4}\rangle$ with $z_{a_\alpha}+z_{b_\alpha}=1$ for $\alpha=1,2,\cdots,N$, while grey ones do not meet this requirement. Black and grey lines represent the intra-dimer $J_0$ and inter-dimer $J_1$ couplings, respectively. }
\label{fig:schematic}
\end{figure}

\begin{figure}
\includegraphics[width=\linewidth]{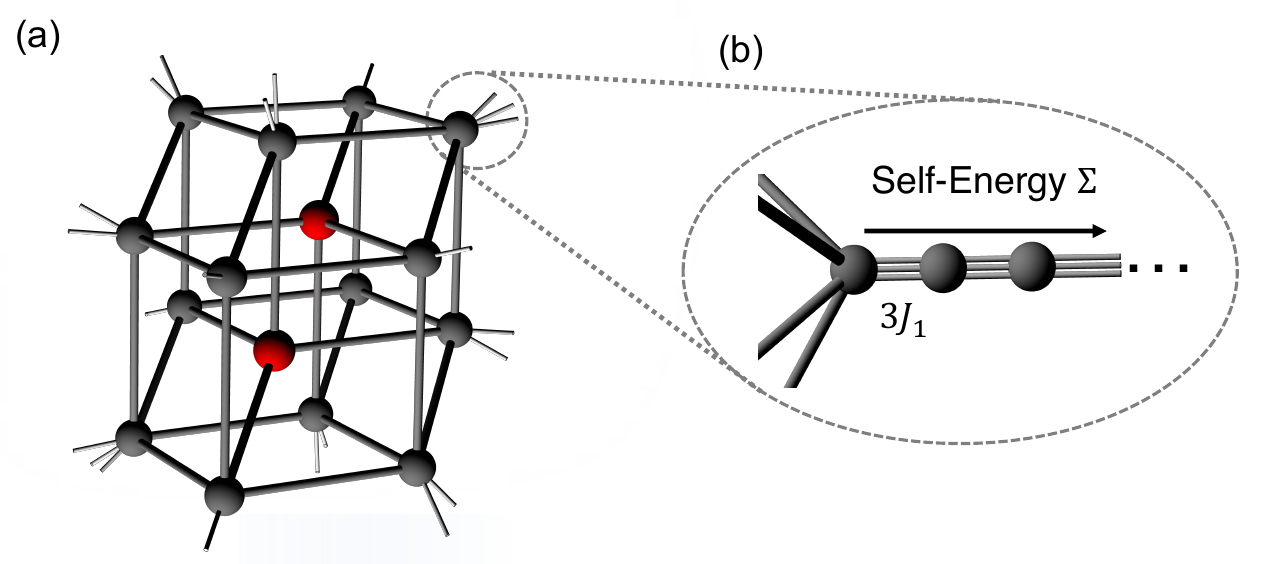}
\caption{(a) The hypercube of the model we discuss in Fig.~\ref{fig:schematic}. For each vertex, we show the coupling with the thermalization region. 
(b) From the perspective of a vertex in the hypercube, the thermal environment can be considered as a semi-infinite chain, represented by the self-energy $\Sigma$. Thus, a set of semi-infinite chains lead to an approximate decay rate of the Hilbert hypercube for a Fock state in the Hilbert hypercube directly connecting to the thermalization region.
}
\label{fig:schematic2}
\end{figure}

From the viewpoint of Hilbert space, each Fock state $\left\vert p\right\rangle=\left\vert z_{a_1}z_{b_1}z_{a_2}z_{b_2}\cdots z_{a_N}z_{b_N}\right\rangle$ is represented by a site, where $z_{a_\alpha}$ and $z_{b_\alpha}$ are qubits $a_\alpha$ and $b_\alpha$ in their ground and excited states, respectively. The connection between two Fock states or sites implies their coupling, describing the particle motion from one qubit to another. 
For the system with four coupled dimers as shown in  Fig.~\ref{fig:schematic}(a), the Hilbert-space adjacency graph can be represented by a robust hypercube connecting to the Hilbert thermalization region as shown in Fig.~\ref{fig:schematic}(b) and  (c). It is noteworthy that the hypercube represents the adjacency graph of the Fock states $\vert p_{\mathrm{H}} \rangle$ with only one excited site in each real-space dimer, i.e. $z_{a_\alpha}+z_{b_\alpha}=1$ for each $\alpha$.

Thus, we can rewrite the Hamiltonian in Hilbert space as:
\begin{equation}
    \begin{split}
        \bm{H} \ \ &= \bm{H}_\mathrm{H} + \bm{H}_\mathrm{T}  + \bm{H}_\mathrm{HT}, \ \ \mathrm{with} \\
        \bm{H}_\mathrm{H} \ &=  \sum_{p_\mathrm{H},q_\mathrm{H}} J_{p_\mathrm{H} q_\mathrm{H}} \vert p_\mathrm{H} \rangle \langle q_\mathrm{H} \vert + \mathrm{h.c.}, \\
        \bm{H}_\mathrm{T} \ &= \sum_{p_\mathrm{T},q_\mathrm{T}}  J_{p_\mathrm{T} q_\mathrm{T}}   \vert p_\mathrm{T} \rangle \langle q_\mathrm{T} \vert  +\mathrm{h.c.} ,\\
        \bm{H}_\mathrm{HT} &=  \sum_{p_\mathrm{H},q_\mathrm{T}}   J_{p_\mathrm{H} q_\mathrm{T}}  \vert p_\mathrm{H} \rangle \langle q_\mathrm{T} \vert  +\mathrm{h.c.} ,
    \end{split}
\end{equation}
where $\bm{H}_\mathrm{H,T}$ denote the single-particle Hamiltonians for the Hilbert hypercube region and for the Hilbert thermalization region, respectively. The coupling coefficient is given by $J_{p, q}=\langle p \vert \bm{H}\vert  q \rangle$. The hoppings between the Hilbert hypercube and thermalization regions are represented by $\bm{H}_\mathrm{HT}$, where indices ``$p,q$'' and ``$p_\mathrm{H/T},q_\mathrm{H/T}$'' stand for the Fock product states in the full Hilbert space and in the hypercube/thermalization subspace, respectively. 

With weak inter-dimer couplings ($J_1$) where each dimer connects to two others, the symmetry of the Hilbert hypercube is broken and the hypercube is connected to other regions. However, there are still two unique collective states at opposite corners of the hypercube that do not directly connect to other regimes. If the system is initially prepared in one of these collective states (for example $\vert C\rangle=\vert 010101\cdots\rangle$ or $\vert C^\prime\rangle=\vert 101010\cdots\rangle$ in quasi 1D comb model), it exhibits slow thermalization dynamics. The Hilbert hypercube has dimension $\mathcal{D}_{\mathrm{H}}=2^N$, while the thermal region has dimension $\mathcal{D}_{\mathrm{T}}=\mathcal{D}-\mathcal{D}_\mathrm{H}$, where $\mathcal{D}=C(2N,N)$ is the total dimension of the Hilbert space. As the system size increases, the thermal region becomes much larger than the Hilbert hypercube region and can be approximated as a thermal bath. Under these conditions, the hopping from the Hilbert hypercube to the thermal region can be considered as radiative decay, distinguishing from the $PXP$ model~\cite{PhysRevB.105.245137}.

For a better understanding of the construction process, we use the 1D SSH model with $2N$ sites in Fig.~\ref{fig:SSH_num_spectrum}(a) as an example to give the explicit form of $\bm{H}_\mathrm{H}$ and $\bm{H}_{\mathrm{HT}}$ which will be used in the following calculation. $\bm{H}_{\mathrm{H}}$ is related to the adjacency matrix of an N-dimension hypercube \begin{equation}
\bm{H}_{\mathrm{H}}=J_0\sum_{\alpha=1}^N \hat{X}_\alpha,
\label{eq:hypercube_hamiltonian_r2}
\end{equation}
which has the same form as the decoupled Hamiltonian in Eq.~(\ref{eq:decouple_hamiltonian}), 
implying the intra-dimer hoppings between two Fock states belonging to the hypercube. For a Fock state $\vert p_{\mathrm{H}}\rangle$ in the hypercube, there is only one particle in each dimer, e.g. $\vert 101010\cdots\rangle\leftrightarrow\vert 011010\cdots\rangle$. Besides, the hopping between the hypercube and the thermal region is represented by $\bm{H}_{\mathrm{HT}}$ in the form of
\begin{equation}
\begin{aligned}
\bm{H}_{\mathrm{HT}}=J_1\sum_{\alpha=1}^{N-1} \Big(&\hat{a}_{\alpha}^+\hat{a}_{\alpha}^-\hat{b}_{\alpha}^+\hat{a}_{\alpha+1}^-\hat{b}_{\alpha+1}^-\hat{b}_{\alpha+1}^+
+\\&\hat{a}_{\alpha}^-\hat{a}_{\alpha}^+\hat{b}_{\alpha}^-\hat{a}_{\alpha+1}^+\hat{b}_{\alpha+1}^+\hat{b}_{\alpha+1}^-+\mathrm{h.c.} \Big),
\end{aligned}
\label{eq:ht_hamiltonian_r2}
\end{equation}
which refers to the inter-dimer hoppings between a hypercube state and a state in the thermalization subspace, e.g. $\vert 101010\cdots\rangle\leftrightarrow\vert 110010\cdots\rangle$. 

We can determine the ratio of the hopping summation between $\bm{H}_{\mathrm{H}}$ and $\bm{H}_\mathrm{HT}$ for the 1D SSH case as follows:\begin{equation}\left(\frac{\Theta}{\Gamma}\right)_{1 \mathrm{D}}=\frac{ N 2^{N} J_0}{\left(N-1\right) 2^{N} J_1} \stackrel{N \rightarrow \infty}{\longrightarrow} \frac{J_0}{J_1},\label{equation:ratio1D}\end{equation}which demonstrates the strength of the hypercube’s connection to the rest of the Hilbert space. Here we define $\Theta = \left(1/2\right)\sum_{p_{\mathrm{H}},q_{\mathrm{H}}}\langle p_{\mathrm{H}}\vert H_{\mathrm{H}}\vert q_{\mathrm{H}}\rangle$ and $\Gamma = \left(1/2\right)\sum_{p_{\mathrm{H}},q_{\mathrm{T}}}\langle p_{\mathrm{T}}\vert H_{\mathrm{HT}}\vert q_{\mathrm{H}}\rangle$. As dissipation is caused by the hopping from the hypercube to the thermal environment (and additionally, the Hilbert space is always too large to consider reflection), the ratio directly influences the fidelity of the scar state and the dynamics of entanglement entropy. The ratio will converge to a finite number as the size of the SSH chain increases, indicating that the scarring phenomenon will occur even in the thermodynamic limit. In Sec.~\ref{sec:dimer_num}, we examine the 1D SSH-chain and quasi-one-dimensional comb lattice cases. We utilize next-next-nearest neighboring coupling to enhance the scarring behavior, which can be viewed as a kind of phase tuning for suppressing the jump from the hypercube to the thermal environment. Furthermore, to verify that this type of scarring extends to the thermodynamic limit, we will conduct a scaling analysis in Sec.~\ref{sec:dimer_num}.C by considering the 2D case.

\subsection{Hypercube decay approximation}
Here, our goal is to estimate the overlap spectrum between special Fock states and eigenstates using a toy model with hypercube decay approximation (HDA). To depict the Hilbert space scarring phenomenon, we employ the non-equilibrium Green’s function approach~\cite{kadanoff2018quantum}. The Green’s function of the Hilbert hypercube is expressed as~\cite{ying2013effect}\begin{equation}\bm{G}(E) =  \Big(  (E+i0^+)\bm{I} - \bm{H}_\mathrm{H} - \bm{\Sigma}(E)   \Big)^{-1},\end{equation}where $\bm{\Sigma}$ is the self-energy that represents the effect of Hilbert thermalization region. By calculating the continuous spectral density of states after the approximation, one can obtain a reasonable estimation of the overlap between the Fock state and the eigenstates in the original discrete spectrum. 
As defined in Eq.~(\ref{eq:hypercube_hamiltonian_r2}), $H_{\mathrm{H}}$ determines the dimension of the non-equilibrium Green's function $G\left(E\right)$, implying that the full Hilbert-space dimension $\mathcal{D}=C(2N,N)$ can be reduced to $\mathcal{D}_H=2^N$.
In general, our model is applicable to other large-scale systems, where the thermalized region is much larger than the hypercube and the connections between them are relatively weak. 

In detail, we can express the self-energy as $\bm{\Sigma}=\sum_{p_\mathrm{H} }\bm{\sigma}_{p_\mathrm{H}}$.  Here, $\bm{\sigma}_{p_\mathrm{H}}(E)=\bm{V}_{p_\mathrm{H}}^\dag \bm{G}\left(E\right)  \bm{V}_{p_\mathrm{H}}$ with $\bm{V}_{p_\mathrm{H}}= \gamma_{p_\mathrm{H}}\vert {p_\mathrm{H}}\rangle\langle {p_\mathrm{H}}\vert $ is the coupling between $\vert p_{\mathrm{H}}\rangle$ and the thermal region. To emulate the radiative decay, we assume that the state ${\vert p_\mathrm{H}\rangle}$ in the Hilbert hypercube is connected to a semi-infinite chain with a hopping strength $\gamma_{p_\mathrm{H}} =   \sum_{q_\mathrm{T}}\langle p_\mathrm{H} \vert\bm{H}  \vert q_\mathrm{T}\rangle =\sum_{q_\mathrm{T}}\langle p_\mathrm{H} \vert\bm{H_{\mathrm{HT}}}  \vert q_\mathrm{T}\rangle$. The hopping strength is the summation of the direct hopping connection from a specific hypercube Fock state to the thermal environment, as shown in Fig.~\ref{fig:schematic2}. As a result, we can obtain the self-energy of state $\vert p_{\mathrm{H}}\rangle$ by solving the self-consistent Dyson equation\begin{equation}\bm{\sigma}_{p_\mathrm{H}}(E) = \bm{V}_{p_{\mathrm{H}}}^\dag \Big( E\bm{I} - \bm{H}_{\mathrm{H}}  -  \bm{\sigma}_{p_\mathrm{H}}(E)  \Big)^{-1} \bm{V}_{p_\mathrm{H}}   .\end{equation}
Then, we have the spectral function (density of states) as
\begin{equation}
    \bm{A}(E) =  -\frac{1}{\pi} \mathrm{Im}\bm{G}(E).
\end{equation}
The local density of states for state $\vert {p_\mathrm{H}}\rangle$ in the Hilbert hypercube is given by
\begin{equation}
A_{p_\mathrm{H}} (E)   =  \langle {p_\mathrm{H}} \vert \bm{A}(E) \vert {p_\mathrm{H}}  \rangle ,
\end{equation}
which implies that Fock state $\vert p_\mathrm{H} \rangle$ overlaps with the continuous `eigenstates'.

\begin{figure}
\includegraphics[width=\linewidth]{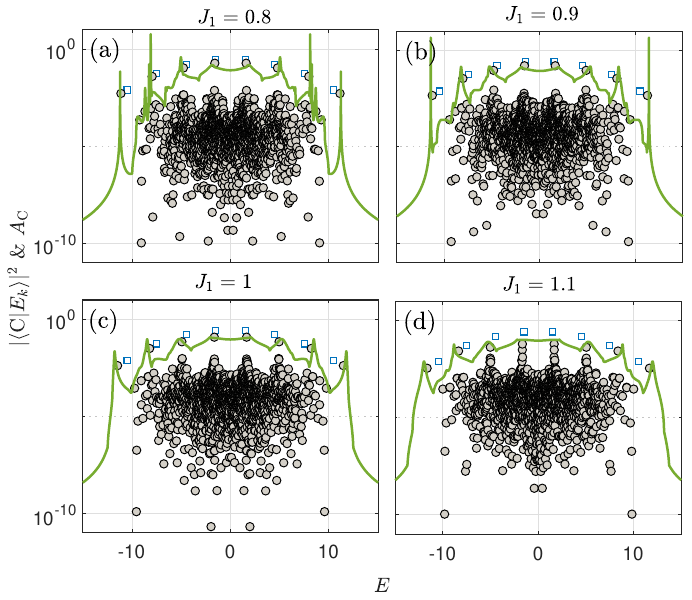}
\caption{
Comparison of overlaps between the collective state $\vert C\rangle$ and eigenstates in the comb lattice for the exact result (black dots) and HDA (green solid curve), respectively. Additionally, the blue squares represent the special eigenstates of a decoupled hypercube ($J_1 = 0$). 
The intra-dimer coupling is fixed at $J_0=1.5$. The inter-dimer coupling in different panels is given by (a) $J_1=0.8$, (b) $0.9$, (c) $1$, and (d) $1.1$, respectively. Here we set the dimer numbers N = 7. }
\label{fig:comparsion}
\end{figure}

We present a comparison between the exact computational results and the toy model using HDA for a comb lattice with $N=7$ in Fig.~\ref{fig:comparsion}. We will discuss the numerical details for the comb lattice in Sec.~\ref{sec:dimer_num}. Our model successfully reproduces the exact result near the zero-energy regime, both for the tower heights and energy. However, this model overestimates the tower heights at high energies. As the strength of the inter-dimer coupling $J_1$ approaches the intra-dimer coupling $J_0$, the overestimation of tower heights gradually becomes weaker. 

\section{Quantum scarring in systems consisting of dimers}\label{sec:dimer_num}

We consider a set of dimers with appropriate inter-dimer coupling $J_1$. There is only one coupling link between two neighbor dimers. In this section, we numerically demonstrate a one-dimensional SSH chain, quasi-one-dimensional comb lattice, and random dimer cluster.

\subsection{Su-Schrieffer–Heeger Chain with perturbation}

\begin{figure}
\includegraphics[width=\linewidth]{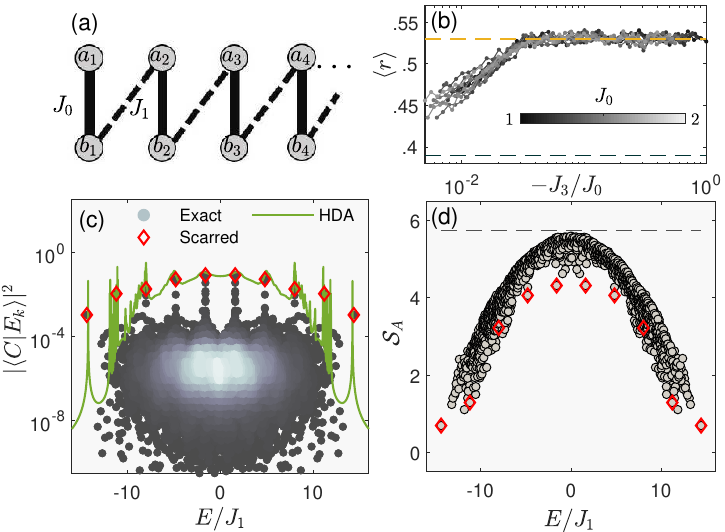}
\caption{
(a) A SSH chain schematic diagram with a set of dimers for the spin-1/2 $XY$ model. The intra- and inter-dimer couplings  are marked by solid and dashed lines, respectively. $a_{m},b_m$ with $m=1,2,3,\cdots$ denote the two sites in a single dimer.
(b) The energy level spacing ratio $\langle r \rangle$ of the SSH chain  as a function of the next-next-nearest-neighbor coupling strength $J_3$ for intra-dimer coupling strength $J_0$ from $1$ to $2$ with a step of $0.1$. 
(c) Scattering plot of the overlap of many-body eigenstates with the special collective state as a function of energy for the 1D SSH chain.
(d) Half-chain bipartite EE as a function of eigenstates for the 1D SSH chain. 
The dimer number in (b-d)  is $N=9$ with $9$ particles. The coupling strengths in (c,d) are given by $J_0/J_1=1.5$ and $J_3/J_1 \approx -0.162$.
}
\label{fig:SSH_num_spectrum}
\end{figure}

\begin{figure}
\includegraphics[width=0.9\linewidth]{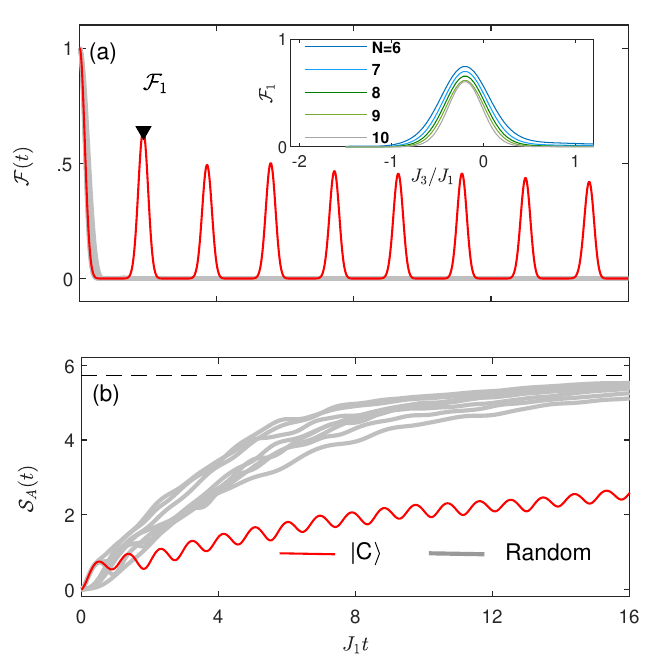}
\caption{ 
(a) Comparison of the dynamics of the fidelity for the collective state $\vert C\rangle$ and $10$ randomly chosen Fock states for the 1D SSH chain system. Inset shows the height of the first fidelity revival $\mathcal{F}_1$ as a function of next-next-nearest-neighbor coupling $J_3$ for systems with different dimer numbers from $N=6$ to $9$.
(b) Comparison of the dynamics of the bipartite EE for the collective state $\vert C\rangle$ and ten randomly chosen Fock states for the 1D SSH chain system. In panels (a,b), the number of dimers is $N=9$, consisting of $9$ particles. The coupling coefficient $J_0/J_1 = 1.6$ and $J_3/J_1 = -0.18$.
}
\label{fig:ssh_num_dynamics}
\end{figure}

\begin{figure}
\includegraphics[width=0.9\linewidth]{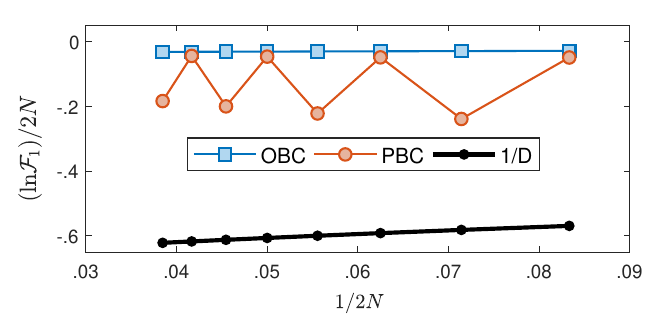}
\caption{The logarithmic fidelity density as a function of the inverse system size $1/2N$ in an SSH chain for the collective initial state $\vert C\rangle$ with OBC and PBC, respectively. The black curve represents the scaling of the inverse Hilbert space dimension, implying the scaling of a typical thermalizing state.}
\label{fig:ssh_num_scaling}
\end{figure}

Here, we numerically study the quantum thermalization properties of the scarring states in the one-dimensional  SSH lattice. The clean SSH Hamiltonian is given by  Hamiltonian in Eq.~(\ref{eq:dimer_hamiltonian}) with 
\begin{equation}
\begin{split}
\beta &=\alpha+1, \ \delta^\prime = 1 ,  \\ 
\delta &=\delta^{\prime\prime}=\delta^{\prime\prime\prime}=0.
\end{split}
\end{equation}
The clean spin-$1/2$ $XY$ SSH chain is integrable. We apply a long-range interaction term, as a perturbation, to break its integrability.
Thus, the Hamiltonian is $\hat{H}_\mathrm{SSH}=H_0+H_3$ with 
\begin{equation}
    \hat{H}_3 = J_3 \sum_\alpha \Big(\hat{a}^+_\alpha \hat{b}^-_{\alpha+1}  +  \hat{b}^+_\alpha \hat{a}^-_{\alpha+2}  + \mathrm{h.c.}  \Big),
\end{equation}
where $J_3$ is the strength of the next-next-nearest-neighbor coupling.  Figure~\ref{fig:SSH_num_spectrum} (a) shows the configuration of the non-integrable SSH Hamiltonian. Such a perturbed SSH chain emerges to be thermalized and its level-spacing statistics agree with the Wigner-Dyson distribution. We numerically obtain the level-spacing ratio $\langle r \rangle$ as a function of the strength of the next-next-nearest coupling $J_3$ for various strengths of $J_1$ and a fixed system size $N=9$, as shown in Fig.~\ref{fig:SSH_num_spectrum} (b). The energy level statistics imply that the non-integrability emerges at the perturbation strength $J_3 > 0.06J_0$. We note that such a non-integrable threshold of $J_3$ is not robust for different system sizes. In general, a larger system has a smaller threshold value of $J_3$.

Now, we focus on the scarred eigenstates.  At first, we plot the overlap between the collective state $\vert C\rangle$ and all eigenstates $\vert E_k\rangle$ for a dimer number of $N=9$ (i.e. $18$ qubits and $9$ particles)  with the coupling coefficients given by $J_0/J_1 = 1.6$ and $J_3/J_1 \approx -0.162$ , as shown in Fig.~\ref{fig:SSH_num_spectrum} (c). Differing from most eigenstates, a set of special states have a strong overlap with the collective states, located at the peaks of towers. These towers have an equal energy spacing, also fitted by a quadratic function for the non-integrable SSH system (see Appendix.~\ref{App:delta E}). We also plot the spectral function of state $\vert C\rangle$ estimated by the HDA. The tower energies in the HDA greatly agree with the exact simulation, while the tower heights away from zero energy are inaccurate.

Furthermore, we calculate the bipartite entanglement entropy. 
The von Neumann expression of the entanglement entropy (EE) is written as
\begin{equation}
\mathcal{S}_A = - \mathrm{Tr}_A  \left[ {\rho_A}\mathrm{ln}{\rho_A} \right] , 
\end{equation}
where $\rho_A = \mathrm{Tr}_B(\vert \phi\rangle\langle\phi\vert )$ is the reduced density matrix of subsystem ``A'' since the system is equally divided by two subsystems ``A'' and ``B''. 
We compute the EE for the one-dimensional SSH chain in Fig.~\ref{fig:SSH_num_spectrum} (a) for the same parameters.  Fig.~\ref{fig:SSH_num_spectrum} (d) shows the eigenstates bipartite EE. It can be observed that the states in the bulk of the entropy spectrum show larger Entanglement Entropy following the volume law. The eigenstate EE of near-zero energy is close to the Page value $\mathcal{S}=  N \mathrm{ln}2 -1/2$. Some special states have relatively smaller EE, manifesting the existence of scarred eigenstates embedded in the thermal energy spectrum. 


Next, we study the fidelity and EE dynamics of different initial product states.
The fidelity is given by
\begin{equation}
\mathcal{F}(t) = {\vert \langle\Psi(0)\vert \Psi(t)\rangle\vert }^2 ,
\end{equation}
where $\vert \Psi(t)\rangle=\mathrm{exp}[-i\hat{H} t]\vert \Psi(0)\rangle$ is the general time-dependant wavefunction.
As shown in Fig.~\ref{fig:ssh_num_dynamics} (a), the fidelity dynamics of a special collective state $\vert C\rangle$ exhibits revival phenomenon with a period of $\Delta T \approx 2\pi/\Delta E$, mainly contributed by the overlap towers in the energy spectrum. These ultra-small thermalized dynamics are distinguished from rapid thermalization for other randomly chosen initial states.  We also compare the time evolution of the EE for the ordinary thermalizing states with the special collective states in Fig.~\ref{fig:ssh_num_dynamics} (b). The EE of the special collective states slowly grows with slight oscillations. Differently, the EE dynamics of the ordinary thermalizing states rapidly increase until its value is close to the Page value.  

In Fig.~\ref{fig:ssh_num_dynamics} (c), we also plot the fidelity at its first revival $\mathcal{F}_1=\mathcal{F}(t_1)$ as a function of $J_3$ for different systems from $N=6$ to $10$. We find that the peak value of $\mathcal{F}_1$ appears on $J_3\approx -0.2 J_0$, while the peak value gradually drops, being accompanied by the increase of the system size. 

Further, we plot the logarithmic fidelity density $(\mathrm{ln}\mathcal{F}_1)/2N$ ~\cite{DHTP:2021} as a function inverse system size $1/2N$ in Fig.~\ref{fig:ssh_num_scaling}. For the randomly chosen initial Fock states, its long-time fidelity is $1/\mathcal{D}$. However, for the collective state $\vert C \rangle$, its scaling of logarithmic fidelity density has a small slope around $0.075$, being distinctive from $1.2$ for ideal thermalizing states. The above discussion is under the open boundary condition (OBC).
For the periodic boundary condition (PBC) case, the logarithmic fidelity density shows a zigzag scaling. The  fidelity density for PBC is close to the OBC case for $N\in \mathrm{even}$, but the larger difference for $N\in \mathrm{odd}$. This difference is gradually relieved as the system size increases, as shown in Fig.~\ref{fig:ssh_num_scaling}.

\subsection{Quasi-one-dimensional comb lattice}

\begin{figure}
\includegraphics[width=\linewidth]{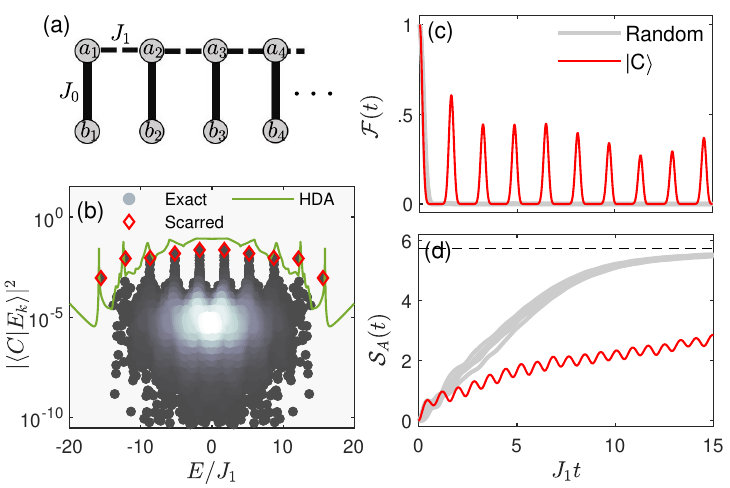}
\caption{(a) Quasi-1D comb lattice consisting of $8$ dimers for the spin-1/2 $XY$ model. The intra-dimer $J_0$ and inter-dimer $J_1$ couplings are denoted by solid and dashed lines, respectively. $a_{m},b_m$ with $m=1,2,3,\cdots$ denote the two sites in a single dimer. (b) Scattering plot of the overlap of many-body eigenstates with the special collective state as a function of eigenenergy for the comb lattice. (c) Comparison of the dynamics of the fidelity with initial Fock states for a collective state $\vert C\rangle$ and ten randomly chosen Fock states. (d) Comparison of the bipartite EE dynamics for the initial Fock states mentioned in (c). In panels (b-d), we consider the many-body comb lattice consisting of $9$ dimers with $9$ particles. The coupling ratio is given by $J_0/J_1=1.6$.}
\label{fig:comb_num}
\end{figure}

\begin{figure}
\includegraphics[width=0.9\linewidth]{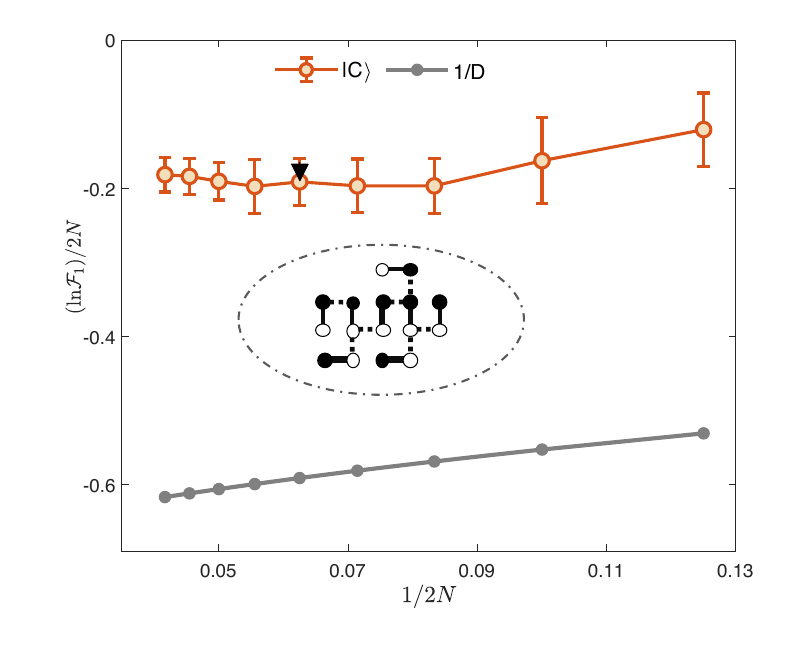}
\caption{The red curve is the logarithmic fidelity density $(\mathrm{ln}\mathcal{F}_1/2N)$ as a function of inverse system size $1/2N$ for the collective initial Fock states $\vert C\rangle$ in random dimer cluster, where $\mathcal{F}_1$ stands for the peak value of the first revival.  Red circles and error bars denote the mean values and standard error for different system sizes, respectively. Each circle and error bar are based on 20 configurations of random dimer clusters.
The black arrow marks the fidelity density of an example for a random dimer cluster consisting of $8$ dimers with $8$ particles(shown in the inset), where the black bullets and circles denote qubits in excited and ground states, respectively, showing one of the collective states $\vert C\rangle$. The grey curve is the scaling of inverse Hilbert space dimension $1/\mathcal{D}$, implying the scaling of typical thermalizing states. In this figure, we consider a coupling ratio of $J_0/J_1$=1.6. 
}
\label{fig:rand_cluster_num}
\end{figure}

Here, we study the quantum  scarring phenomenon of the scarring states in the quasi-one-dimensional comb systems. The origin of the quantum scars is the same as the SSH model, from the robustness of the Hilbert cube. The only difference is in the connection between the cube and the thermal region in Hilbert space. The Hamiltonian of a comb lattice is also given by the Hamiltonian in Eq.~(\ref{eq:dimer_hamiltonian}), but with different parameters of 
\begin{equation}
\begin{split}
    \beta & = \alpha +1,  \ \delta^{\prime\prime}=1, \\
    \delta & = \delta^{\prime}= \delta^{\prime\prime\prime}=0.
\end{split}
\end{equation}
The clean comb model is non-integrable and thus another perturbation term is not necessary.

We consider a comb lattice consisting of $9$ dimers with the configuration in Fig.~\ref{fig:comb_num}~(a).  The couplings are given by $J_0=1.6$ and $J_1=1$.
The square overlaps $\vert \langle C \vert  E_k \rangle\vert ^2$ show sharp towers in Fig.~\ref{fig:comb_num}~(b), while the eigenstates are denser in each tower than the above SSH model. The HDA estimation perfectly fits the tower energy. In Fig.~\ref{fig:comb_num}~(c,d), we plot the fidelity and EE dynamics both for a collective state and randomly chosen states. We find that the collective state has strong fidelity revivals and slow growth of EE, in contrast to the near-zero fidelity and rapid growth of EE for randomly chosen initial states. We also note that the variance of those randomly chosen states in EE dynamics is quite small, in particular for late time.

We also note that a previous study of the comb system with offshoots of random lengths was shown to have ``compact" localized states, as well as the quantum scarring phenomenon based on a model with density-density interactions~\cite{PhysRevResearch.2.043267}. The origin of such scarring is from the single-particle compact localized state in a unit consisting of three backbone sites, the mechanism of which is distinct from the study here.

\subsection{Random dimer cluster}

Furthermore, we consider random configurations of the dimer cluster. The Hamiltonian is given by Eq.~(\ref{eq:dimer_hamiltonian}), restricting that each site has at most two hopping connections to other dimers, and there is only one coupling link between two neighbor dimers. As for choosing the initial state, the phase of paired dimers is $\pi$. The inset of Fig.~\ref{fig:rand_cluster_num} illustrates an example of a random dimer cluster consisting of $8$ dimers. The number of configuration combinations  exponentially rises with the increase in system size. We calculate the scaling of the logarithmic fidelity density of the first revival in Fig.~\ref{fig:rand_cluster_num}. We average over 20 random configurations in each system size. The average logarithmic fidelity density scaling of the collective state implies that such a collective state can survive as the system size increases, with a remarkable difference from the scaling of $1/\mathcal{D}$ for thermalized states.

\section{Hilbert space scars in higher dimensions}\label{sec:2d}

\begin{figure}
\includegraphics[width=\linewidth]{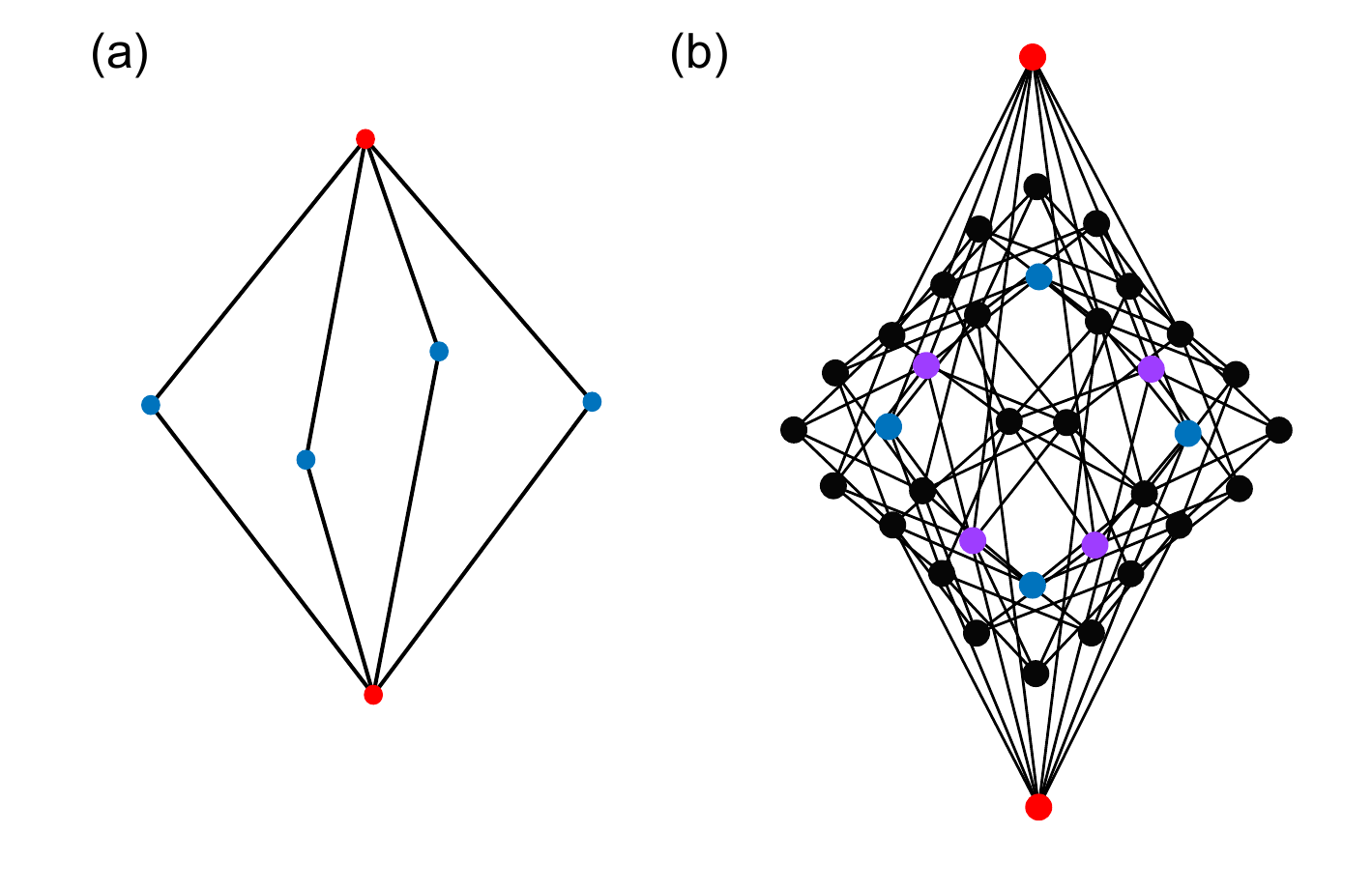}
\caption{The Hilbert hyperpolyhedron with two particles in each tetramer in real space for (a) a tetramer and (b) two tetramers, respectively. Red bullets denote a pair of second collective states  $\left\{\vert C_\times\rangle, \vert C^\prime_\times\rangle \right\}$. Blue bullets denote the first collective states $\left\{ \vert C_{\vert\vert}  \rangle, \vert C^\prime_{\vert\vert}  \rangle, \vert C_= \rangle,\vert C^\prime_= \rangle\right\}$, respectively. Purple bullets stand for those mixed Fock states of first and second classes, which also have no direct connection to the Hilbert thermalization region.
}
\label{fig:2d_hilbert}
\end{figure}

\begin{figure}
\includegraphics[width=1\linewidth]{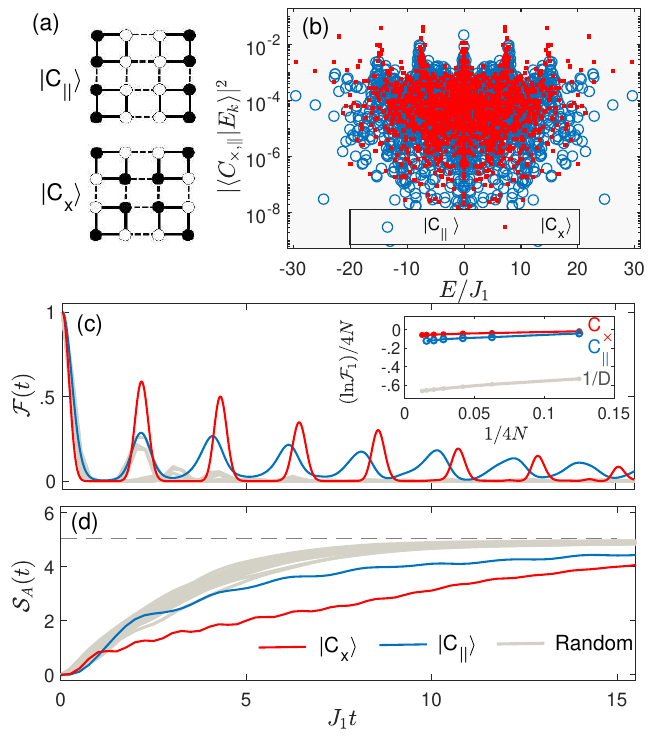}
\caption{(a) Two classes of the collective Fock states $\vert C_{\vert\vert }\rangle$ and $\vert C_{\times}\rangle$ for the 2D SSH model consisting of four tetramers. Solid and dashed lines denote the couplings $J_0$ and $J_1$, respectively. The black bullets and circles denote qubits in excited and ground states, respectively. (b) Overlaps between the collective Fock states $\vert C_{\vert\vert }\rangle$ or $\vert C_{\times}\rangle$ and eigenstates.
(c,d) show the fidelity and bipartite EE dynamics for $\vert C_{\vert\vert }\rangle$, $\vert C_{\times}\rangle$, and twenty randomly chosen Fock states, respectively. 
In panels (b-d), we consider $N_x=N_y=2$, i.e. four tetramers in total with $8$ particles. The coupling ratio is given by $J_0/J_1=2.5$. Inset of (c) shows the scaling of logarithmic fidelity density $(\mathrm{ln}\mathcal{F}_1/4N)$ for state $\vert C_\times\rangle$, $\vert C_{\vert\vert} \rangle$, and inverse Hilbert space dimension $1/\mathcal{D}$ from the TDVP with a bond dimension of $256$ and a time step of  $0.02$.}
\label{fig:2d_num}
\end{figure}

In this section, we study the quantum many-body Hilbert space scars in higher dimensions, in which the unit of a lattice is not a dimer anymore. For example, the unit is a tetramer and octamer in 2D and 3D SSH many-body models, respectively.

Similar to dimer systems in low dimensions, the intra- and inter-tetramer couplings are denoted by $J_0$ and $J_1$, respectively. Each tetramer consists of four qubits, allowing us to consider a 2D many-body SSH lattice composed of a set of tetramers. For a clean 2D SSH lattice in the spin-$1/2$ $XY$ model, the system is non-integrable. In Hilbert space, a special subset is distinct from other Hilbert thermalization regions, particularly for $J_0 > J_1$, called the Hilbert hyperpolyhedron, which resembles the hypercube in low-dimensional systems. The Hilbert hyperpolyhedron is spanned by the Fock product states with two particles in each tetramer, and is more intricate than the low-dimensional cases composed of dimers, as depicted in Fig.~\ref{fig:2d_hilbert} for one and two tetramers. 
More generally, the definition of the Hilbert hyperpolyhedron is the adjacency graph of the Fock states $\vert p_{\mathrm{H}} \rangle$, where each basic unit is half-filled. Here, the basic unit denotes the dimer for the 1D/quasi-1D case, the tetramer for the 2D case, and the octamer for the 3D case.

In the Hilbert hyperpolyhedron, a series of classes of Fock product states are special with collective features between adjacent tetramers. Among them, two classes of collective states, $\vert C_{\vert \vert }\rangle$ and $\vert C_\times\rangle$, exhibit remarkable scarred dynamics. The first class of collective states $\vert C_{\vert \vert }\rangle$ (or $\vert C_{=}\rangle$) is essentially associated with the 1D SSH case along the $x$ (or $y$) axis while it is simply a replica of a 1D SSH chain along the $y$(or $x$) axis, as illustrated in the upper panel of Fig.~\ref{fig:2d_num} (a). The second class, $\vert C_\times\rangle$, is a true 2D collective state where two diagonal sites in a tetramer are occupied by two particles and two adjacent tetramers have a $\pi$ phase difference, as illustrated in the lower panel of Fig.~\ref{fig:2d_num} (a). These two classes of states in the Hilbert hyperpolyhedron lack a hopping connection to the thermal region, while decoupled states are not limited to those. Other decoupled states are combinations of these two classes.

In the energy spectrum from eigenstates, both $\vert C_{\vert\vert}  \rangle$ and $\vert C_\times\rangle$ have high overlap with special eigenstates, resulting in a set of equally spacing towers, as shown in Fig.~\ref{fig:2d_num}~(b). 
The difference is that the overlap of $\vert C_\times\rangle$ is overall higher than $\vert C_{\vert\vert} \rangle$ with lanky towers. This allows that the fidelity dynamics of $\vert C_\times\rangle$ show higher revivals, as shown in Fig.~\ref{fig:2d_num}~(c). On the contrary,  other randomly chosen states quickly thermalize and lose their initial information. Furthermore, the remarkable differences of those initial states are distinct in bipartite EE dynamics in Fig.~\ref{fig:2d_num}~(d). In addition, we note the oscillation period of $\vert C_{\vert\vert} \rangle$ in EE dynamics is double of $\vert C_\times\rangle$, which is caused by the configuration difference of these two initial Fock states and the cut of two subsystems. 

Differently, from the low-dimensional system, the scarred state in the 2D case is hard to be verified by the scaling from the exact numerics as the exponential expansion of the Hilbert space dimension exists. However, referring to the low-dimensional case in Secs.~\ref{sec:model} and \ref{sec:dimer_num}, we also can approximately give the ratio of the hopping summation over between $\bm{H}_\mathrm{H}$ and $\bm{H}_\mathrm{HT}$ for the 2D case by 
\begin{equation}
   \left(\frac{\Theta}{\Gamma}\right)_{2\mathrm{D}} = \frac{(4/3) N_xN_y 6^{N_x N_y} J_0 }{(2N_x N_y - N_x - N_y) 6^{N_x N_y} J_1}  \xrightarrow{N_x,N_y \rightarrow \infty  }  \frac{2J_0}{3J_1}, 
\end{equation}
where $N_{x,y}$ denote the tetramer numbers along $x$ and $y$ axises, respectively. Then, the total tetramer number is given by $N=N_xN_y$.
Thus, this distribution behavior of the collective states is distinct from the others beyond the hyperpolyhedron region, even in the thermodynamic limit. Also, in the inset of Fig.~\ref{fig:2d_num}~(c), for different system sizes,  we calculate the logarithmic fidelity density $(\mathrm{ln}\mathcal{F}_1/4N)$ for state $\vert C_\times\rangle$ shown in Fig.~\ref{fig:2d_num}~(a) by using time-dependent variational principle (TDVP)~\cite{kramer2005geometry,ho2019periodic,yang2022exploring} and its scaling of inverse system size $1/4N$ suggests states $\vert C_\times\rangle$ and $\vert C_\times^\prime\rangle$ can survive in the thermodynamic limit. However, the scaling curve of state $\vert C_{\vert\vert} \rangle$ is more tilted than $\vert C_\times\rangle$. Thus, it is hard to conclude its existence in the thermodynamic limit and further study is needed.

\begin{figure}
\includegraphics[width=\linewidth]{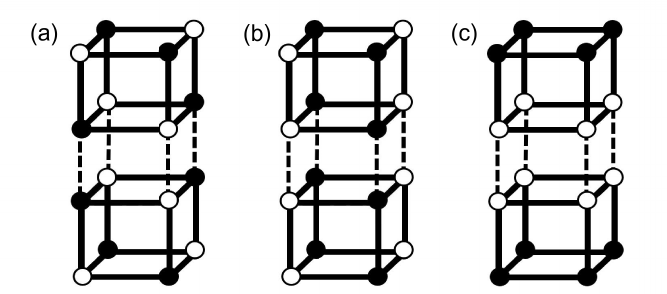}
\caption{  Three classes of collective Fock states for 3D SSH lattice consisting of two octamers. Solid and dashed lines denote the couplings $J_0$ and $J_1$, respectively. The black bullets and circles denote qubits in excited and ground states, respectively.}
\label{fig:3d}
\end{figure}

The mechanism of Hilbert space scars from 1D to 2D is reminiscent of similar scars in three dimensions. Naively, we first construct two classes of special collective states in the x-y plane, as mentioned in the 2D case, and then replicate them to each plane along the z-axis. Using two octamers as an example, we show the configuration of the 3D case $\vert C_{\times}\rangle$ and $\vert C_{\vert \vert }\rangle$ in Fig.~\ref{fig:3d} (b) and (c), respectively. To seek more remarkable scarred dynamics, we should introduce the initial state configuration shown in Fig.~\ref{fig:3d} (a), denoted by $\vert C_*\rangle$. The key feature of $\vert C_*\rangle$ is that we can see an SSH chain when we look along any straight line in any direction. This provides a viable solution for creating quantum scars originating from hypercubes in arbitrary dimensions.  As a rough feasibility verification, we can also approximately give the ratio of the hopping summation over between $\bm{H}_\mathrm{H}$ and $\bm{H}_\mathrm{HT}$ for the 3D case by 
\begin{equation}
\begin{aligned}
   \left(\frac{\Theta}{\Gamma}\right)_{3\mathrm{D}} &= \frac{(24/7) N_xN_yN_z70^{N_x N_y N_z} J_0 }{2\times(3N_x N_y N_z - N_x N_y - N_y N_z - N_z N_x) 70^{N_x N_y N_z} J_1} \\& \qquad\qquad\qquad\qquad\qquad\qquad\xrightarrow{N_x,N_y,N_z \rightarrow \infty  }  \frac{4J_0}{7J_1}. 
   \end{aligned}
\end{equation}

Similar to the case in 1D and 2D, that ratio finally will be a finite number even under an infinite scale. More generally, we can derive the ratio in $M$-dimensional SSH case 
\begin{equation}
\left(\frac{\Theta}{\Gamma}\right)_{M \mathrm{D} } \xrightarrow{\mathrm{size}\rightarrow\infty} \frac{2^{M-1} J_0}{\left(2^M-1\right) J_1}.
\end{equation}
The result will converge to a finite number whatever how large M is, implying that the Hilbert space quantum scar in high dimensions also originated from the hyperpolyhedron. 


\section{discussions}

In summary, we systematically study a distinct class of QMBS based on a hard-core Bose-Hubbard model in arbitrary dimensions. In the simplest form, this model is consisting of a set of dimers and it is different from the conventional constrained model, typified by the $PXP$ model. We find that, in general, the origin of this kind of scar is caused by a robust subspace with specific hypercube configurations surrounded by a thermalization region with slightly weak couplings in Hilbert space, thus it is also called Hilbert space quantum scar. Within this hypercube subspace, there exist special Fock states, called collective states, which lack direct coupling to the Hilbert thermalization region. These collective states exhibit periodic fidelity revivals and show an almost linear growth of the bipartite entanglement entropy as a function of time, unlike other randomly chosen Fock states. Additionally, some of the special scarred eigenstates have a significant overlap with these collective states, revealing equal-spacing towers in the energy spectrum. Moreover, these special eigenstates show a lower bipartite entanglement entropy compared to other eigenstates. As the system size increases, the volume of the Hilbert thermalization region exponentially expands, leading us to assume that it acts as a thermal bath. 

To verify the assumption, our analysis shows that the radiative decay of the hypercube into the thermalization region can be accurately described using the non-equilibrium Green’s function approach in this toy model, with the predicted overlap spectrum agreeing with the results from exact numerical simulations. Through our investigations of the eigenstates and dynamics of the collective Fock states, we have demonstrated various dimer lattice configurations, including the 1D SSH, comb lattice, and random dimer cluster. Furthermore, we explore Hilbert quantum scars in higher dimensions, where the higher-dimensional lattice is consisting of tetramers or octamers for two and three dimensions, respectively. Those higher-dimensional Hilbert space quantum scars have the same origin as the low-dimensional case, while more classes of special Fock states with remarkable fidelity revivals can be found. We also calculate the scaling results of the 1D SSH, random cluster, and 2D SSH models. The scaling of those quantum many-body scar states implies that the scarring phenomenon can survive as the system size increases, with a remarkable difference from the scaling behavior for thermalized states. 

This model and the lattice configurations are experimentally friendly since a number of platforms--such as superconducting circuit-QED systems, cold atoms, and ion trapping systems--can simulate the hard-core Bose-Hubbard Hamiltonian in a straightforward way. In addition, the initial state is a Fock product state, which is easy to be prepared in many experimental platforms. For instance, in the superconducting platform, the coupling strength can be modified by tuning the detuning of a coupler qubit connecting to two nearest-neighbor qubits, resulting in the observation of this kind of QMBS in 1D~\cite{zhang2022many} and 2D~ \cite{yao2022observation} SSH configurations. Furthermore, realization in such platforms can improve the coherent time of specific many-body states, which highlights the potential for utilizing Hilbert space quantum scars in quantum sensing and quantum metrology~\cite{Dooley:2021, dooley2023entanglement, desaules2022extensive}.

\section*{Acknowledgment}
We thank Z. Papić and J.-Y. Desaules for their valuable discussion. We acknowledge support from the National Key R\&D Program of China (Grant No. 2022YFA1404203).

%

\appendix

\section{Numerical fitting of the equal spacing of eigenstates $\Delta E$}\label{App:delta E}

\begin{figure}[htb]
\includegraphics[width=\linewidth]{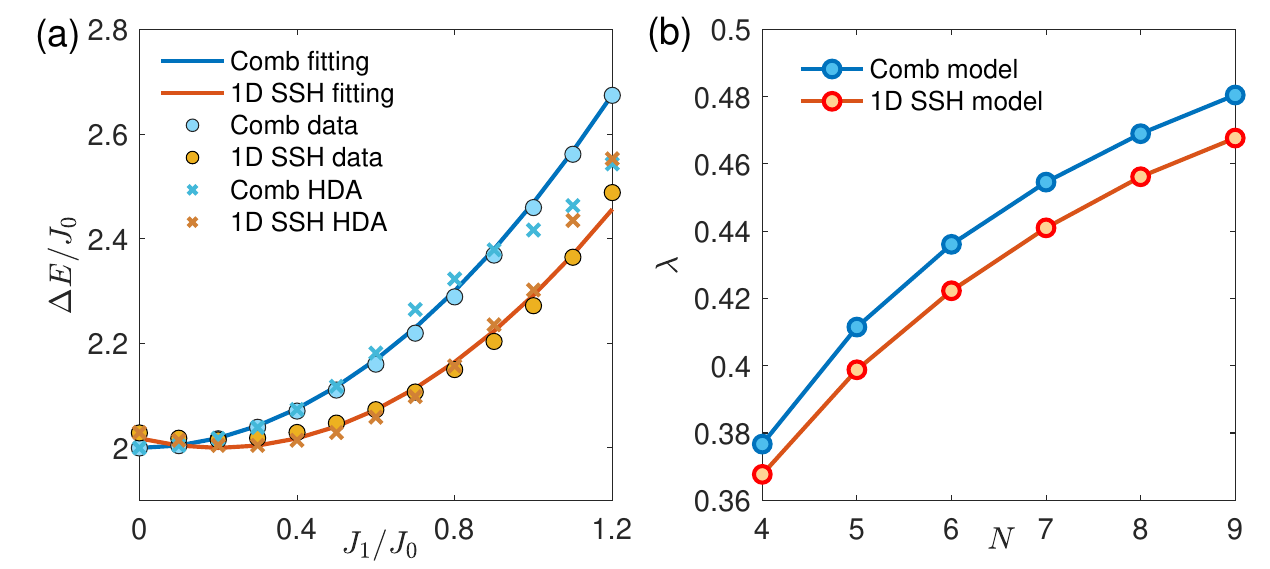}
\caption{
(a) Polynomial fitting of the equal spacing of eigenstates $\Delta E$. The dots show the numerical results of $\Delta E$ while the lines show the polynomial fitting curve. Besides, the crosses represent the results calculated by the HDA method. The fitting process considers the data with $J_0/J_1$ in the interval of [0, 1.2], and the number of dimers is set as N = 8. (b) Curve fitting parameter $\lambda$ according to system size. In both panels (a) and (b) Orange and blue lines(dots, crosses) show the results of the 1D SSH model and the comb lattice respectively.
}
\label{afig:delta_e}
\end{figure}

We've discussed in the main text that $\Delta E$ of the non-integrable systems such as the 1D SSH model and comb lattice can be fitted by quadratic functions with a size-dependent parameter $\lambda$. For comb lattice, we can fit it by $\Delta E/J_0 \approx \lambda(J_1/J_0)^2+2$ which is mentioned before. Meanwhile, we should consider the effect of the next-next-nearest coupling $J_3$ in the non-integrable 1D SSH model. Here we choose $J_3/J_0 = -0.2$ as an example. We find that the fitting function approximately changes to $\Delta E/J_0 \approx \lambda(J_1/J_0+J_3/J_0)^2+2$. Both of their fitting results are shown in  Fig.~\ref{afig:delta_e}(a). If $J_1/J_0$ is small, approximately in the interval between 0 and 1, the quadratic function fits very well. While the data points outsides this interval hold a tendency to gradually deviate from the fitted curve because the correlation between hypercube and thermal bath becomes stronger when $J_1/J_0$ increases. Similarly, the HDA method becomes invalid to predict $\Delta E$ well as the inter-dimer coupling $J_1$ increases.

Furthermore, we show how fitting parameter $\lambda$ changes according to the system size. In both the comb model and the 1D SSH model, the size-dependent parameter increase as the dimer number change from $N=4$ to $N=9$, while the growth rate is gradually slowing down.

\end{document}